\documentclass{PoS}
\usepackage{amsmath,amssymb}

\title{Chiral symmetry breaking and monopoles}

\ShortTitle{Chiral symmetry breaking and monopoles}

\author{Adriano Di Giacomo\\
        University of Pisa, Department of Physics and INFN, Sezione di Pisa, Largo B. Pontecorvo 3, Pisa, 56127, Italy\\
        E-mail: \email{digiaco@df.unipi.it}}

\author{\speaker{Masayasu Hasegawa}\\
        Joint Institute for Nuclear Research, Bogoliubov Laboratory of Theoretical Physics, Dubna, Moscow, 141980, Russia\\
        E-mail: \email{hasegawa@theor.jinr.ru}}

\author{Fabrizio Pucci \\ 
        Universite Libre de Bruxelles, Service 3BIO, Avenure F.Roosevelt 1050, Brussels, Belgium \\ 
        E-mail: \email{fapucci@ulb.ac.be}}

\abstract{To understand the relation between the chiral symmetry breaking and monopoles, the chiral condensate which is the order parameter of the chiral symmetry breaking is calculated in the $\overline{\mbox{MS}}$ scheme at 2 [GeV]. First, we add one pair of monopoles, varying the monopole charges $m_{c}$ from zero to four, to SU(3) quenched configurations by a monopole creation operator. The low-lying eigenvalues of the Overlap Dirac operator are computed from the gauge links of the normal configurations and the configurations with additional monopoles. Next, we compare the distributions of the nearest-neighbor spacing of the low-lying eigenvalues with the prediction of the random matrix theory. The low-lying eigenvalues not depending on the scale parameter $\Sigma$ are compared to the prediction of the random matrix theory. The results show the consistency with the random matrix theory. Thus, the additional monopoles do not affect the low-lying eigenvalues. Moreover, we discover that the additional monopoles increase the scale parameter $\Sigma$. We then evaluate the chiral condensate in the $\overline{\mbox{MS}}$ scheme at 2 [GeV] from the scale parameter $\Sigma$ and the renormalization constant $Z_{S}$. The final results clearly show that the chiral condensate linearly decreases by increasing the monopole charges.}

\FullConference{The 8th International Workshop on Chiral Dynamics,\\
		29 June 2015 - 03 July 2015\\
		Pisa, Italy}

\begin{document}

\section{Introduction}

 We want to show that monopoles are closely related to instantons and to chiral symmetry breaking by using the Overlap fermions, which preserve the chiral symmetry in the lattice gauge theory~\cite{Gins1,Neuberger1}. In order to show the relation, we add one pair of one monopole and one anti-monopole by the monopole creation operator, which is developed in ~\cite{DiGiacomo1}.

 In a previous work~\cite{DiGM1} the following results were derived:\\
(i) One pair of a monopole with positive magnetic charge and an anti-monopole with negative magnetic charge creates instantons.\\
(ii) The spectral density of the low-lying eigenvalues of the Overlap Dirac operator increases by increasing the monopoles charge.\\
(iii) The chiral condensate decreases when increasing the monopole charge. (This result was directly computed from the eigenvalues.)
 
 In this study, we generate the normal configurations by the standard heat-bath and over-relaxation methods. The action is the Wilson gauge action. The parameter set of the lattice is $V = 14^{4}$, $\beta = 6.00 \ (a/r_{0}=0.1863)$, and $N_{conf} = O (^{}2\times10^{3})$. We generate configurations with additional monopoles varying the magnetic charges $m_{c}$. We use the same parameter set and standard methods as for the normal configurations. The number of configurations is $N_{conf} = O (^{}2\times10^{3})$ for each value of the magnetic charge. We compute low-lying eigenvalues and eigenvectors of the Overlap Dirac operator from the gauge links of these configurations by solving the eigenvalue problem. (The details are in Ref.~\cite{DiGM1}. We increase the number of the configurations for this study.) [Section~\ref{subsec:Ov_1},~\ref{subsec:Mon_1}]

 Next, in order to investigate effects of the additional monopoles, we compute the actual level spacing and the unfolding level spacing, using the low-lying eigenvalues of the Overlap Dirac operator. We compare the distributions of the nearest-neighbor spacing with the prediction of the random matrix theory~\cite{Edwards1}. We then compare the low-lying eigenvalues without depending on the scale parameter $\Sigma$ with the prediction of the random matrix theory~\cite{Giusti1}. Our results are compatible with the results in the literature, and are consistent with the predictions of the random matrix theory. Therefore, we evaluate the scale parameter $\Sigma$ by two different ways~\cite{Edwards1,Giusti1}. We discover that the scale parameter $\Sigma$ increases by increasing the charge of the monopoles [Section \ref{sec:rmt}].

 Finally, we compute the renormalization constant $Z_{S}$ from the pseudo-scalar masses which are computed from the correlation functions. We precisely evaluate the chiral condensate from the scale parameter $\Sigma$ and the renormalization constant $Z_{S}$ in the $\overline{\mbox{MS}}$ scheme at 2 [GeV]~\cite{Wennekers1}. The chiral condensate in the continuum limit is $\Sigma^{\overline{MS}} = -\{285 (4) \ [\mbox{MeV}]\}^{3} \ (r_{0} = 0.5 \ [\mbox{fm}])$, which is calculated from normal configurations, and interpolated by using five different values of the lattice spacing. Here, the final results clearly show that the chiral condensate in the $\overline{\mbox{MS}}$ scheme decreases by increasing the monopole charges [Section \ref{sec:chi_co}]. 

\subsection{Overlap Dirac operator}\label{subsec:Ov_1}

 The massless Overlap Dirac operator $D(\rho)$ is defined from the massless Wilson Dirac operator $D_{W}(\rho)$. The massless Wilson Dirac operator $D_{W}(\rho)$ is defined with the (negative mass) parameter $\rho$ = 1.4 as follows:
\begin{equation}
D_{W}(\rho) = D_{W} - \frac{\rho}{a}
\end{equation}
The lattice spacing is $a$. Using the massless Hermitian Wilson Dirac operator $H_{W}(\rho)$, the massless Overlap Dirac operator $D(\rho)$ is transformed as follows: \begin{equation}
D(\rho) = \frac{\rho}{a} \left\{ 1 +  \gamma_{5}\epsilon(H_{W}(\rho)) \right\}
\end{equation}
$\epsilon$ is the sign function.

We solve the eigenvalue problem by the Arnoldi method using the ARPACK subroutines, and save $O(60-80)$ pairs of low-lying eigenvalues and eigenvectors of the Overlap Dirac operator. The improved eigenvalue (being projected onto the imaginary axis) is a pure imaginary number, close to the values in the continuum limit~\cite{Capitani1}. These eigenvalues always appear in pairs of opposite sign on the imaginary axis, that is as $\pm i\lambda^{imp}$. In this study we use the improved ``positive'' eigenvalue $\lambda^{imp}$, and omit the superscript from now on in this report.

\subsection{Monopoles}\label{subsec:Mon_1}

 We consider monopoles  with positive charges $m_{c} = 0, +1, +2, +3, +4$ and anti-monopoles with negative charges $-m_{c} = 0, -1, -2, -3, -4$. We add one pair of monopole and anti-monopole of magnetic charge $\pm m_{c}$ ranging from zero to four by the monopole creation operator~\cite{DiGiacomo1}. The total magnetic charge of the additional monopoles is zero. Hereafter, $m_{c}$ indicates that both the positive and negative magnetic charges $\pm m_{c}$ are added. We locate the monopole and the anti-monopole at a given distance.
 We have found that this monopole creation operator only produces long monopole loops in configurations. The monopole density linearly increases with the monopole charge. We have shown that the additional monopoles do not affect the detection of the zero modes of the Overlap fermion by comparing the distributions of the topological charges with our predictions~\cite{DiGM1}. 

\section{Random matrix theory}\label{sec:rmt}

 The random matrix theory predicts universally the spectral density distributions of the Dirac operator. The microscopic spectral density is provided in Ref.~\cite{Verbaarschot2}. In the $\epsilon$ regime, the distribution functions of the $k$-th smallest eigenvalues of the Dirac operator are derived in Ref.~\cite{Damgaard1, Damgaard2}. 

\subsection{The nearest-neighbor spacing}

 The nearest-neighbor spacing distribution $P(s)$ is used to check short range fluctuations in the eigenvalues~\cite{Guhr1}. The Gaussian ensembles can be classified (according to Wigner) into three different classes: the Gaussian orthogonal ensemble, the Gaussian unitary ensemble, and the Gaussian symplectic ensemble. The distribution function of the Gaussian unitary ensemble is given by
\begin{equation}
P(s) = \frac{32}{\pi^{2}}s^{2}\exp\left(-\frac{4s^{2}}{\pi}\right)\label{eq:dis_uni_1}.
\end{equation}
\begin{figure}[htbp]
  \begin{center}
    \includegraphics[width=70mm]{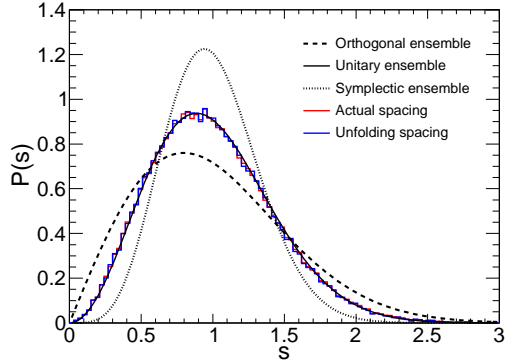}
  \end{center}
  \caption{A comparison of distributions of the nearest-neighbor spacing with the random matrix theory. The predictions are indicated by the dashed, solid, and dotted lines. The distributions computed from the eigenvalues are drawn by the red and blue lines. The normal configurations are used. The lattice is $V = 14^{4}, \ \beta = 6.00$.}
  \label{fig:level_sp_or_1}
\end{figure}
\begin{figure}[htbp]
  \begin{center}
    \includegraphics[width=150mm]{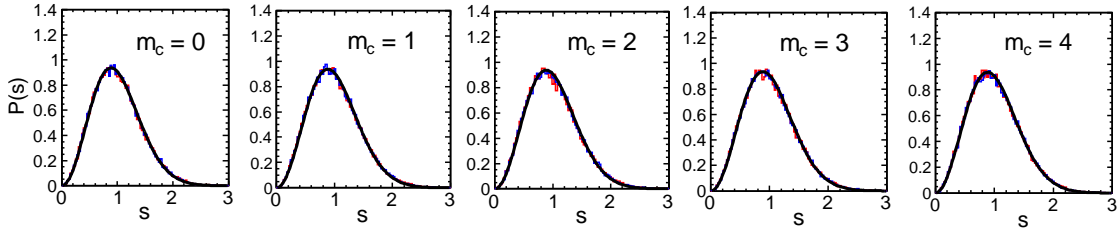}
  \end{center}
  \caption{Comparisons of distributions of the nearest-neighbor spacing. The configurations with additional monopoles are used. The monopole charges $m_{c}$ are varying from zero to four. The prediction of the random matrix theory is drawn by the solid line. Our results are indicated by the red and blue lines. The lattice is $V = 14^{4}, \ \beta = 6.00$.}\label{fig:level_sp_add_1}
\end{figure}

 We suppose that the distribution of the nearest-neighbor spacing in this study is consistent with the unitary symmetry ensemble, because an evidence was shown by R. G. Edwards and his collaborators in Ref.~\cite{Edwards1}. Therefore, we examine the consistency of the distributions with the random matrix theory. First, we compute the distribution from actual spacing as follows: $P_{act}(s) \equiv \frac{s_{i}}{\langle s_{i}\rangle}, \ s_{i} = \lambda_{i+1} - \lambda_{i}, \ (i > 0)$. $\langle\cdots\rangle$ is the average value on the number of configurations. The distribution is indicated by the red lines in Fig.~\ref{fig:level_sp_or_1} and \ref{fig:level_sp_add_1}. Next, we compute the unfolding level spacing by the configuration unfolding process in Ref.~\cite{Guhr2}. We determine the smooth part of the cumulative spectral function $N_{ave}(\lambda)$, by fitting the polynomial function (d = 5) to the mean values of eigenvalues $\langle\lambda_{i}\rangle$. The unfolding level spacing is computed as follows: $P_{unf}(s) \equiv N(\lambda_{i + 1}) -  N(\lambda_{i})$. The distribution is indicated by the blue lines in Fig.~\ref{fig:level_sp_or_1} and \ref{fig:level_sp_add_1}. We normalize the distributions to unity.

 Our data clearly obey the distribution of the Gaussian unitary ensemble $P(s)$~(\ref{eq:dis_uni_1}) as shown in Fig.~\ref{fig:level_sp_or_1} and \ref{fig:level_sp_add_1}. Moreover, the ensemble class is compatible with the result of Ref.~\cite{Edwards1}.

\subsection{Scale-independence test}

 In the $\epsilon$-regime of QCD, the random matrix theory predicts the distribution of the $k$-th scaled eigenvalue $z_{k}$ of the Dirac operator in each topological charge sector $|Q|$~\cite{Damgaard1,Damgaard2}. The $k$-th scaled eigenvalue obeys the relation
\begin{equation} 
z_{k, |Q|}^{RMT} = \Sigma V \lambda_{k, |Q|}.\label{eq:rmt_eq_1} 
\end{equation}
\begin{figure}[htbp]
  \begin{center}
    \includegraphics[width=125mm]{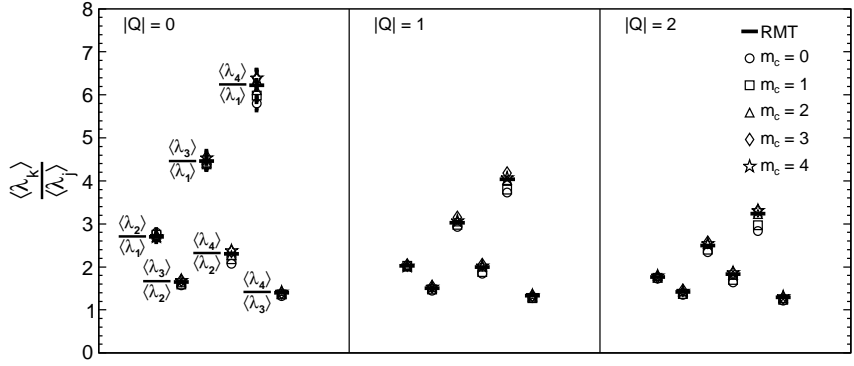}
  \end{center}
  \caption{The comparison of the scale independent observable with the random matrix theory in each topological charge sector $|Q|$. The configurations with additional monopoles, $V = 14^{4}$, $\beta = 6.00$ are used. The monopole charges vary from zero to four.}\label{fig:scale_test_Add_mon_1}
\end{figure}
We compute the observable $\frac{\langle \lambda_{k} \rangle}{\langle \lambda_{j}\rangle}$, which is independent of the scale parameter $\Sigma$, in order to check the low-lying spectrum of the Overlap Dirac operator~\cite{Giusti1}. We compare the results computed from configurations with additional monopoles, with the prediction of the random matrix theory. The results are consistent with the random matrix theory as shown in Fig.~\ref{fig:scale_test_Add_mon_1}. The spectra are not affected by the additional monopoles.
\begin{figure}[htbp]
  \begin{center}
    \includegraphics[width=120mm]{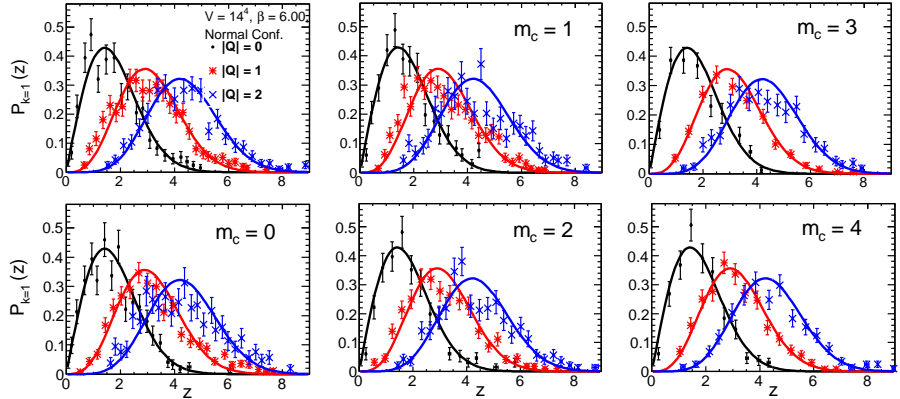}
  \end{center}
  \caption{The comparisons of the distributions of the first eigenvalues in each topological charges sector $|Q| = 0, 1, 2$ with the random matrix theory. The smooth curves indicate the predictions from the random matrix theory. Our lattice data are scaled using the fitted results of $\Sigma^{(2)}$ in each $|Q|$.}\label{fig:scale_2m0-4ch8d_1}
\end{figure}
\subsection{Scale parameter $\Sigma$}

 The distribution functions of the first eigenvalues ($k = 1$) in each topological charge sector are derived from the random matrix theory as follows~\cite{Edwards1,Damgaard1,Damgaard2}:
\begin{equation}\label{eq:fdis_rmt_1}
P_{k=1}(z)= \begin{cases}
\frac{z}{2}\exp\left(-\frac{z^{2}}{4} \right) & |Q|= 0, \\
\frac{z}{2}I_{2}(z)\exp\left(-\frac{z^{2}}{4} \right) & |Q| = 1, \\
\frac{z}{2}\{I_{2}^{2}(z) - I_{1}(z)I_{3}(z)\}\exp\left(-\frac{z^{2}}{4} \right) & |Q| = 2.
\end{cases}
\end{equation}
$I_{\nu}(z)$ is the modified Bessel function of the first kind, $\nu \ > \ 0$. We evaluate the scale parameter $\Sigma$ in two different ways. First, we analytically compute $\Sigma^{(1)}$ from Eq.~(\ref{eq:rmt_eq_1}) and~(\ref{eq:fdis_rmt_1}). Second, we create functions with one free parameter ($\Sigma^{(2)}$) from Eq.~(\ref{eq:fdis_rmt_1}). We determine $\Sigma^{(2)}$ by fitting the functions to the observed distributions of the lowest eigenvalues in each topological sector $|Q|$~\cite{Edwards1}. We compare the distributions of the lowest eigenvalues in each topological charge sector with~(\ref{eq:fdis_rmt_1}) as shown in Fig.~\ref{fig:scale_2m0-4ch8d_1}. The results are consistent. The mean values of $\Sigma^{(1)}$ and $\Sigma^{(2)}$ in each topological charge sector are consistent. We average the values of the three topological charge sectors. The scale parameter increases with the value of the magnetic charge as indicated in Table~\ref{tb:condensate_new}. 

\section{The chiral condensate}\label{sec:chi_co}

 We evaluate the regulated chiral condensate $\langle \bar{\psi} \psi \rangle^{\overline{MS}}$ by the modified minimal subtraction ($\overline{\mbox{MS}}$) scheme at 2 [GeV] based on Ref.~\cite{Wennekers1}. There is a relation, in $N_{f} = 3$ quark flavors, between the scale parameter and the chiral condensate as follows:
\begin{equation}
\Sigma = - \langle\bar{\psi}\psi\rangle
\end{equation}
We use this relation, and evaluate the chiral condensate in quenched QCD from $\Sigma$ and $Z_{S}$. 

 The renormalization group invariant (RGI) scale parameter $\hat{\Sigma}$ is calculated from the renormalization constant $Z_{S}$. We compute $Z_{S}$ from the relation $Z_{S} = \frac{1}{Z_{M}}$. $Z_{M}$ is computed by converting the bare quark to the RGI mass as follows:
\begin{equation}
\hat{Z}_{M}(g_{0}) = \frac{U_{m}}{m_{q}r_{0}}\Bigr|_{(m_{ps}r_{0})^{2} = x_{ref}}
\end{equation}
$m_{q}r_{0}$ is the bare quark (valence quark) mass. We set the reference quark mass $x_{ref} = 1.5736$, therefore, $U_{m} = 0.181(6)$~\cite{Hernandez1}. We measure the pseudo-scalar mass to get $m_{q}r_{0}$ at the reference mass, using the lattice of $V = 14^{3}\times28$, $\beta = 6.00$, and $N_{conf} = 400$ in each measurement. Here, the pseudo-scalar mass is measured by a subtraction of the scalar correlation from the pseudo-scalar correlation~\cite{DeGrand1}. Next, the RGI scale parameter $\hat{\Sigma}$ is 
\begin{equation}
\hat{\Sigma} = \frac{\Sigma}{\hat{Z}_{S}}.
\end{equation}
$\Sigma$ is computed from $V = 14^{4}$, and $\hat{Z}_{S}$ is computed from $V = 14^{3}\times28$. Finally, we convert $\hat{\Sigma}$ to the $\overline{\mbox{MS}}$ scheme at 2 [GeV] as follows: $\hat{\Sigma}^{\overline{MS}} (2 \ [\mbox{GeV}]) = \hat{\Sigma}/z, \ \ z = 0.72076$. The values of $\hat{Z}_{S}$, $\hat{\Sigma}$, and the chiral condensates are summarized in Table~\ref{tb:condensate_new}. 
\begin{table*}
\small
\caption{The final results. $r_{0}^{3}\Sigma^{(1)}$ and $r_{0}^{3}\Sigma^{(2)}$ are mean values of three topological charge sectors. The chiral condensates are computed from $r_{0}^{3}\hat{\Sigma}_{\overline{MS}}^{(1)}$, $\left(r_{0} = 0.5 \ [\mbox{fm}]\right)$. N. C. stands for the normal configurations.}\label{tb:condensate_new}
{\renewcommand{\arraystretch}{1.0}
\begin{tabular*}{\textwidth}{c @{\extracolsep{\fill}}cccccc} \hline\hline
$m_{c}$  & $r_{0}^{3}\Sigma^{(1)}$ & $r_{0}^{3}\Sigma^{(2)}$ & $\hat{Z}_{S}$ & $r_{0}^{3}\hat{\Sigma}_{\overline{MS}}^{(1)}$ & $\langle\bar{\psi}\psi\rangle^{\overline{MS}}$ [GeV$^{3}$]& $\{\langle\bar{\psi}\psi\rangle^{\overline{MS}}\ [\mbox{MeV}^{3}]\}^{1/3}$ \\ \hline
N. C. &  0.263(6) & 0.264(7)  & 1.02(3) &  0.373(16) & -0.0230(10) & -284(4)\\
0     & 0.259(6)  &0.262(7)   & 0.98(3) &   0.353(15) & -0.0217(9) & -279(4) \\
1     & 0.284(7)  &0.286(8)   & 1.00(3) &   0.396(17) & -0.0243(10) & -290(4) \\
2     & 0.353(11) &0.362(9)   & 0.97(3) &   0.48(2) & -0.0293(14) & -308(5) \\
3     & 0.406(11) &0.403(11)  & 0.98(3) &   0.55(2) & -0.0340(15) & -324(5) \\
4     & 0.417(11) &0.418(11)  & 0.97(3) &   0.56(2) & -0.0347(15) & -326(5) \\\hline\hline
\end{tabular*}
}
\end{table*}

 Lastly, we briefly show the result of the chiral condensate at the continuum limit. We set one physical lattice volume $V/r_{0}^{4} = 49.96$, and generate normal configurations by five different parameters of the lattice spacing ($\beta = $ 5.812, 5.904, 5.989, 6.000, 6.068). The number of configurations is $O (1\times10^{3})$ in each $\beta$. We then compute the RGI scale parameters $\hat{\Sigma}$ in each $\beta$, from the scale parameters $\Sigma^{(1)}$ and $\hat{Z}_{S}$. $\hat{Z}_{S}$ is analytically computed using an interpolating function in each $\beta$~\cite{Wennekers1}. Finally, we interpolate the five RGI scale parameters to the continuum limit, and obtain a final result $r_{0}^{3}\hat{\Sigma} = 0.271(12)$. The chiral condensate at the continuum limit ($r_{0}$ = 0.5 [fm]) is 
\begin{equation}
\langle \bar{\psi} \psi \rangle^{\overline{MS}} \ \ (2 \  [\mbox{GeV}]) \ \ = \ \ -2.31 (11)\times10^{-2} \ \ \ [\mbox{GeV}^{3}] \ \  =  \ \ -\{ \ 285 (4) \ \ \ [\mbox{MeV}] \ \}^{3}.\label{eq:chi_condens_mev}
\end{equation}
This result is consistent with Ref.~\cite{Wennekers1,Giusti2,Giusti3}.

\subsection{Chiral symmetry breaking and monopoles}

\begin{figure}[htbp]
  \begin{center}
    \includegraphics[width=145mm]{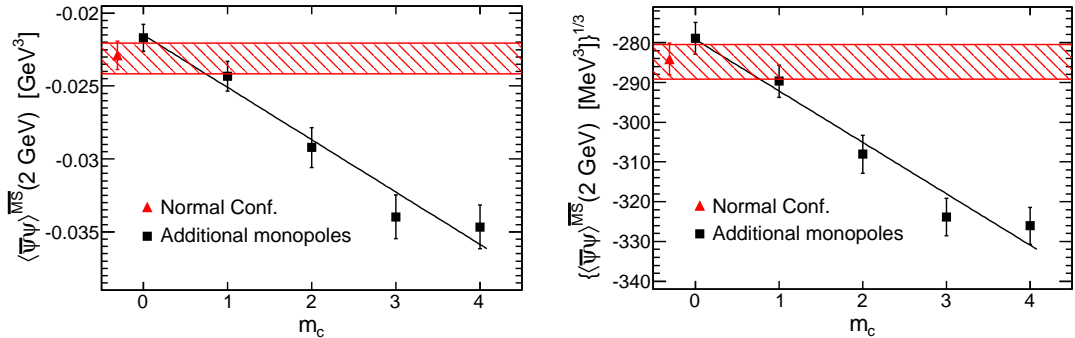}
  \end{center}
  \caption{The chiral condensate in [GeV$^{3}$] and [MeV] unit. The results computed using the normal configurations are indicated by the red symbols. The range of values of the chiral condensate in the continuum limit is indicated by the two horizontal bands.}\label{fig:Chiral_add_or_conti_mev_1}
\end{figure}

 In Fig.~\ref{fig:Chiral_add_or_conti_mev_1} we show the results obtained by fitting our final data with the following linear functions: $\langle\bar{\psi}\psi\rangle^{\overline{MS}} = A\cdot m_{c} + B$ \ (left figure), and $\{\langle\bar{\psi}\psi\rangle^{\overline{MS}}\}^{1/3} = C\cdot m_{c} + D$ \ (right figure). The fitted parameters (and the $\chi^2/d.o.f.$) are
\begin{align}
A & = -0.0036(4), \ \ \ B = -0.0215(8), \ \ \ \chi^{2}/d.o.f. = 2.7/3.0 \ \ (\mbox {left figure}),\\
C & = -12.9(1.4), \ \ \ D = -279(3), \ \ \ \ \ \ \ \ \chi^{2}/d.o.f. = 3.4/3.0 \ \ (\mbox{right figure}). 
\end{align}
These results indicate that the chiral condensate linearly decreases with the value of the monopole charge.
\section{Summary and Conclusion}

 We computed the distributions of the nearest-neighbor spacing. We performed the scale-independence test. We confirmed that monopoles do not affect the low-lying eigenvalues of the Overlap Dirac operator. The monopoles increase the scale parameter $\Sigma$. Finally, we have shown that the chiral condensate linearly decreases by increasing the monopole charges. This is an evidence that monopoles are related to the chiral symmetry breaking. The chiral symmetry breaking is induced by the monopole condensation in QCD.

\section{Acknowledgments}

 The simulations have been performed on SX-ACE, SX-8, SX-9, and PC clusters at RCNP and CMC at the University of Osaka, and SR16000 at YITP at the University of Kyoto. We really appreciate their technical supports and the computational time. M. H. thanks R. G. Edwards for  useful discussion and advice at the international workshop on Chiral Dynamics 2015 in Pisa.

\end{document}